\documentclass[aps,pre,twocolumn,groupedaddress,showkeys,showpacs]{revtex4-1}
\usepackage{amsmath,amssymb}
\usepackage[utf8]{inputenc}
\usepackage{graphicx}
\usepackage{float}
\usepackage{color}

\begin{document}

\title{Extracting work from a single reservoir in the non-Markovian underdamped regime}

\author{Oscar Paredes-Altuve}
\affiliation{Centro de F\'isica, Instituto Venezolano de Investigaciones Cient\'ificas, 21827, Caracas, 1020 A, Venezuela.}
\author{Ernesto Medina}s

\affiliation{Centro de F\'isica, Instituto Venezolano de Investigaciones Cient\'ificas, 21827, Caracas, 1020 A, Venezuela}
\affiliation{Yachay Tech, School of Physical Sciences \& Nanotechnology, 100119-Urcuqu\'i, Ecuador.}
\author{Pedro J. Colmenares}
\affiliation{Grupo de Qu\'imica Te\'orica: Qu\'imicof\'isica de Fluidos y Fen\'omenos Interfaciales (QUIFFIS), Departamento de Qu\'imica,
Facultad de Ciencias, Universidad de los Andes, M\'erida, Venezuela.}

\begin{abstract}
We derive optimal-work finite time protocols for a colloidal particle in a harmonic well in the general non-Markovian underdamped regime in contact
with a single reservoir. Optimal work protocols with and without measurements of position and velocity are shown to be linear in time. 
In order to treat the underdamped regime one must address forcing the particle at the start and at the end of a protocol, conditions which dominate the
short time behaviour of the colloidal particle. We find that for protocols without measurement the least work by an external agent decreases linearly
for forced start-stop conditions while those only forced at starting conditions are quadratic (slower) at short times, while both decrease asymptotically to zero for quais-static processes. When measurements are performed protocols with start-end forcing are still more efficient at short times but can be
overtaken by start-only protocols at a threshold time. Measurement protocols derive work from the reservoir but always below the 
predicted by the Sagawa's generalization of the second law. Velocity measurement protocols are more efficient in deriving work than position measurements.\\

Published in Phys. Rev. E, {\bf 94}, 062111 (2016).

 DOI: 10.1103/PhysRevE.94.062111
\end{abstract}  

\pacs{}
\date{\today}

\maketitle

\section{Introduction} 
Systems described by stochastic thermodynamics are characterized by having observable slow degrees of freedom associated
with `small' objects such as colloidal particles, biopolymers, molecular engines etc, and unobservable fast degrees of freedom associated
with the fluid/reservoir in which the small particle is immersed. The time scale separation between the fast and
the slow degrees of freedom allow for an appropriate thermodynamic description with a well defined temperature\cite{SeifertReview}. 
For the slow degrees of freedom, an ensemble of trajectories can be defined through the distribution of initial states and
the system evolves through the dynamics determined by both external driving forces and the thermal stochastic forces of the
fluid reservoir. For this system, new fluctuation theorems have been developed that are 
applicable far from equilibrium (time dependent driving) and steady state
(constant driving) situations\cite{jarz,crooks}. Driving such systems in a particular manner, so as to achieve a desired
result e.g. a certain amount of work be performed, has been called a {\it protocol}\cite{SeifAbreu,SeifertReview,bauer}. Such a protocol can be subject
to constraints such as specified displacement or time, or required to be optimal, thus allowing the derivation of new laws in this mesoscopic
realm of thermodynamics.

New refined optical/mechanical techniques have allowed the design of protocols performed after a measurement is executed\cite{cantilever} 
on e.g. a colloidal particle. These kind of systems emulate the thought experiment of Maxwell's Demon\cite{rev:max} and 
they are known as {\it feedback close-loop controled system}\cite{Cao}. This approach has been followed by Abreu and Seifert\cite{SeifAbreu} ans Pal {\it et al.}\cite{1bath} 
who propose ways to extract work from a single heat bath. The study of these systems has suggested new or extended fluctuation relations
that evidence a generalization of the second law of thermodynamics that includes the information gained from measurement\cite{Sagawa}. 

In this work we study protocols that optimize the work applied to the colloidal particle both in the absence and presence
of a measurement of the particle's position (a localizing laser) and velocity (doppler effect)\cite{exp3,joubad,Doppler}. Here we contemplate both inertial
and non-Markovian effects for the colloidal particle. Therefore our treatment departs from the Generalized Langevin Equation (GLE), that includes 
a memory in the form of a friction kernel\cite{ColmenaresOlivares}. The GLE was used previously in the study of generalized fluctuations theorems
(FT)\cite{NoMarkov1,NoMarkov2}. 

The analysis of non-Markovian fluctuations described by generalized Langevin equations (GLE) can be done through alternative approaches such as the Fluctuations theorems (FT) of stochastic thermodynamics. They evaluate the probability distribution of functionals, like work, heat and entropy changes, along an ensemble of trajectories with a given well-defined initial distribution\cite{SeifertReview}. In fact, for the problem posed in Eq. (1), Mai and Dhar\cite{NoMarkov1}, Speck and Seifert\cite{NoMarkov2} and Ohkuma and Ohta\cite{ohkuma} determined that for an exponential kernel\cite{berne}, the Jarzynski equality\cite{jarz}, the transient FT\cite{crooks} and Crooks FT\cite{crooks2} are shown to be exact. These results directly validate Berne's exponential model\cite{berne} as a choice for the memory kernel in the GLE. Moreover, their definitions of work, heat and energy change coincides with the ones used in this work. A compilation of recent works in stochastic thermodynamics can be found in Van den Broeck et al.\cite{broeck}.

The structure of the paper is as follows: In Section \ref{dos} we assume the dynamics of the colloidal particle in a harmonic potential as non-markovian, that is, it is inertially driven by  a generalized Langevin equation (GLE). Its associated bivariate Fokker-Planck equation (FPE) is provided. Additionally, we show in this section the expressions for work and heat performed on the brownian particle. The initial state of the system for the three measurement cases considered in this work: position, velocity  and both, position and velocity, are treated in Section \ref{tres}.
In Section \ref{cuatro}, we derive the work performed for the {\it instantaneous} protocol as a reference for performance, while in section \ref{cinco}, we show that the average velocity of the center of the trap obeys an integral equation in order to optimize the average work.  In the latter, we compare a few functional forms for the velocity of the center of the harmonic potential that accomplish this optimal criterion. In Section \ref{seis} we take the optimal protocol for the work function of the previous section and add the information gained by the measurement. Finally in Section \ref{siete} we show a non-optimal but intuitive protocol that offers results comparable to the optimal ones during a specific range of time. We end with the conclusions.

%------------------------------------------------

\section{The Model}
\label{dos}

\subsection{Dynamics}
We treat the simple model consisting of a colloidal particle of mass $m$ under the action of a harmonic potential whose position
is described by the GLE
\begin{equation} \label{GLE}
m \ddot{x}(t) + \int_0^t \Gamma(t-s) \dot{x}(s) ds + \kappa \left( x(t)- \lambda(t)\right)  = R(t),  
\end{equation} 
where $\Gamma(t)$ is the friction kernel, $\kappa$ is the harmonic well spring constant, $\lambda(t)$ is the position of the center
of the harmonic well and $R(t)$ is a homogeneous, stationary, zero mean Gaussian colored noise with correlation function
given by the fluctuation dissipation theorem i.e.  $\langle R(t) R(s) \rangle = 2 k_{\rm B} T ~\Gamma(t-s) $, where $k_{\rm B}$ and $T$, are the
Boltzmann constant and temperature respectively.

The average properties of the particle described by the equation of motion require to know $P(x,v,t)$ associated with the solution 
of Eq. \eqref{GLE}. In order to find the expression for $P(x,v,t)$, we resort to the stochastic Liouville equation\cite{vanK} and Novikov's theorem\cite{novikov}, as described in Ref. \cite{ColmenaresOlivares}:
\begin{equation}\label{FPEG}
\partial_t P(x,v,t) = - \partial_x J_x - \partial_v J_v,
\end{equation} 
where $J_x$ and $J_v$ are the probability currents 
\begin{align*}
J_x &=  v P(x,v,t), \\
J_v &=  \! - \left[ \! \int_0^t \! \frac{ \Gamma(t \! - \! s)}{m} v(s) ds+ \frac{\kappa}{m} (x\! - \! \lambda(t)) \right] P(x,v,t)\\ 
& -\frac{ k_{\rm B} T}{m^2} \frac{\partial P(x,v,t)}{\partial v} \int_0^t \Gamma(t-s) \frac{d \chi_v(t-s)}{dt} ds\\
& -\frac{ k_{\rm B} T}{m^2}\frac{\partial P(x,v,t)}{\partial x}  \int_0^t \Gamma(t-s) \chi_v(t-s) ds, 	
\end{align*} 
and $\chi_v(t)$ is a Green's function obtained from the solution of Eq. \eqref{GLE}. The initial initial conditions for $\chi_v(t)$
are assumed to be  $\chi_v(0)=0$ and $d\chi_v(t)/dt \mid_{t=0} = 1$. It satisfies  the relation
\begin{equation}
\hat{\chi}_v = \frac{ m }{ m k^2 + k \hat{\Gamma}(k) + \kappa },
\end{equation}
where the symbol $~\widehat{\cdot}~$ indicates the Laplace Transform. As described in reference \cite{adelman},
Eq. \eqref{FPEG} shows an extra diffusive term, proportional to $\partial^2 P(x,v,t)/\partial v \partial x$, that disappears in the markovian limit.
From now on we will denote ensemble averages values, with respect to $P(x,v,t)$, with bold fonts.

%------------------------------------------------

\subsection{Thermodynamics}
Work on the system is performed by manipulating $\lambda(t)$, the position of the center of the harmonic well 
(see Eq. (\ref{GLE})). 
The time dependence of this parameter will be known as {\it the protocol}. In order to write down an expression for the work, we will depart from the conservation of energy for a trayectory $x(t)$\cite{booksek}. Conservation of energy along a 
trajectory $x(t)$ dictates
\begin{equation}
d W = d E - d Q.
\end{equation}
where $E$ is the internal energy and $Q$ the exchanged heat. $dW$ represents the work applied to the system and $dQ$ is the heat
transfered to it.   

The Van Kampen Lemma\cite{vanK} shows the equivalence between computing the average over the realizations of noise $R(t)$ distributed according to $P(R(t))$ and computing the average over $P(x,v,t)$. This is essentially because  the average of the density of points in $x,v$ space over realizations of $R(t)$ is the $P(x,v,t)$. This useful principle (as it is easier to analytically perform calculations with $P(x,v,t)$ )  is used in the classic works on heat and work in Stochastic thermodynamics of Sekimoto, and particularly in the references we have cited by the Seifert group.

Averaging with respect to $P(x,v,t)$ and integrating
in time between $0$ and $t_f$, we obtain the ensemble average work performed on the particle as
\begin{equation}\label{Work}
\mathbf{W}= \mathbf{E}(t_f) - \mathbf{E}(0) + \int_0^{t_f}  \frac{\left< d  Q \right>}{dt} dt, 
\end{equation}  
where $t_f$ is a predetermined final time.The energy of the system $E$ is given by the sum of the kinetic and the potential energies
\begin{equation}
E(t) = \frac{m v^2}{2} +\frac{\kappa}{m} \left(x -\lambda(t) \right)^2,
\label{InternalEnergy}
\end{equation} 
and the heat transfered is given by \cite{booksek}
\begin{equation}
\frac{\left<d Q\right> }{dt} = \int \left( J_x  \partial_x E + J_v \partial_v E   \right) dx dv.
\label{HeatProduced}
\end{equation} 
Substituting Eqs. (\ref{InternalEnergy}) and (\ref{HeatProduced}) into Eq. (\ref{Work}), and making use of the equation of motion
for the probability $P(x,v,t)$ (Eq. (\ref{FPEG})), we obtain (see appendix A for details)
\begin{eqnarray} 
\label{work}
\mathbf{W} &=&  \frac{\kappa}{2} \left( \left( \mathbf{x}(t_f) - \lambda(t_f) \right)^2 - \left( \mathbf{x}(0) - \lambda(0) \right)^2 \right)\notag \\
&+& \frac{m}{2} \left(  \mathbf{v}(t_f)^2 - \mathbf{v}(0)^2 \right)\notag \\
&+& \int_0^{t_f} \mathbf{v}(t)    \int_0^{t} \Gamma(t-s) \mathbf{v}(s)  ds dt. 
\end{eqnarray}
The minimum work that can be performed on the system to change its state is bounded by the expression
\begin{equation} \label{2daley}
\mathbf{W} \geq \Delta F,
\end{equation} 
where $F$ is the free energy of equilibrium states with $\lambda(t)$ held constant, and is given by
\begin{equation}
F \equiv - \ln \left( \int \exp(- E / k_{\rm B} T ) dx  \right).
\end{equation} 
The equality in Eq. \eqref{2daley} is achieved in the quasi-static limit  i.e. for sufficiently
slow protocols.

In order to minimize the work performed in Eq. \eqref{work}, one needs to define the initial distribution of the system. In the
case that no initial measurement is performed on the system, we assume an equilibrium distribution, otherwise we must
account for the new information in the initial distribution as discussed in the following section.
 
 %------------------------------------------------
 
 \section{Measurements }\label{tres}
 
Making a measurement of the position and/or the velocity of the particle renders information necessary to infer the state of the system.
In our model, the particle is at equilibrium when the measurement takes place so that, at that time, the velocity and the position
are uncorrelated. The latter implies that measuring one of the variables does not change the distribution of the other. 
Thereby, the strategy of Abreu and Seifert\cite{SeifAbreu} can be used to analyze the underdamped regime.
 
We assume that the measured value of position, $x_m$, is distributed as a Gaussian probability distribution around the 
actual position of the particle $x$
\begin{equation*}
	P(x_m|x)= \frac{1}{\sqrt{2 \pi \Delta_x^2} } e^{-\frac{(x_m - x)^2}{2 \Delta_x^2}},
\end{equation*}
where $\Delta_x^2$ is the error associated with the measurement. 
The distribution of the position at equilibrium $P_{\rm eq}(x)$ is also Gaussian distributed
\begin{equation*}
P_{\rm eq}(x)= \frac{1}{\sqrt{2 \pi \frac{k_{\rm B} T}{\kappa}} } e^{-\frac{(x - \lambda_0)^2}{2 \frac{k_{\rm B} T}{\kappa}}},
\end{equation*} 
around the center of the harmonic well $\lambda_0$. The probability that one makes a measurement at
$x_m$, $P(x_m)$ can be derived from
\begin{eqnarray}\label{measurementxm}
P(x_m) &=& \int P_{\rm eq}(x) P(x_m|x) dx\notag \\
	&=& \frac{1}{\sqrt{2 \pi \left(\Delta_x^2 + \frac{k_{\rm B} T}{\kappa}  \right) } } e^{-\frac{(x - \lambda_0)^2}{2 \left(\Delta_x^2 + \frac{k_{\rm B} T}{\kappa}  \right) }},
\end{eqnarray} 
where we can see that the uncertainty due to the equilibrium distribution is compounded by the error in the measurement.
With this result and Bayes theorem, $P(x|x_m)  P(x_m)= P_{\rm eq}(x) P(x_m|x)$, it is possible to write the probability distribution
for the particle's actual position $x$,  $P(x|x_m)$
\begin{equation}\label{Pxm}
P(x|x_m) = \frac{1}{\sqrt{2 \pi y_x^2  } } e^{-\frac{(x - b_x)^2}{2 y_x^2 }},
\end{equation}
where
\begin{align*}
	y_x^2 &= \frac{k_{\rm B} T \Delta_x^2 }{k_{\rm B} T + \kappa \Delta_x^2},  \\
	b_x  &= \frac{k_{\rm B} T x_m + \Delta_x^2 \lambda_0 \kappa }{k_{\rm B} T + \kappa \Delta_x^2}.
\end{align*} 
The joint probability that the particle is actually at $x$ and has velocity $v$ at the begining of the protocol is $P_i(x,v)=P_{\rm eq}(v) P(x|x_m)$, where 
\begin{equation*}
P_{\rm eq}(v) = \frac{1}{\sqrt{2 \pi \frac{k_{\rm B} T}{m}} } e^{-\frac{v^2}{2 \frac{k_{\rm B} T}{m}}},
\end{equation*} is the velocity distribution in equilibrium. In the same fashion one can obtain the distribution of velocities after performing a measurement  $v_m$
\begin{equation}\label{Pvm}
P(v|v_m)= \frac{1}{\sqrt{2 \pi y_v^2  } } e^{-\frac{(v - b_v)^2}{2 y_v^2 }},
\end{equation} 
where
\begin{align*}
	y_v^2 &= \frac{k_{\rm B} T \Delta_v^2 }{k_{\rm B} T + m \Delta_v^2},  \\
	b_v  &= \frac{k_{\rm B} T v_m  }{k_{\rm B} T + m \Delta_v^2}.
\end{align*} 
In this case, the initial distribution is expressed as $P_i(x,v)=P_{\rm eq}(x) P(v|v_m)$. When one performs both position and velocity measurements, the initial distribution will be given by
\begin{equation}
P_i(x,v|x_m,v_m) = P(x|x_m) P(v|v_m).
\end{equation}

A way to quantify the amount of information gained in a measurement involves the Kullback-Leibler distance\cite{Info} 
$I(\cdot)$, where $\cdot$ is the measured variable that compares the probability distributions after the measurement
with the equilibrium distribution. Thus, we have that information gained measuring position, velocity or both
are given by
\begin{align}
	\overline{I(x_m)} &= \frac{1}{2} \log \left( \frac{k_{\rm B} T}{\kappa \Delta_x^2} +  1  \right), \\
	\overline{I(v_m)} &= \frac{1}{2} \log \left( \frac{k_{\rm B} T}{m \Delta_v^2} +  1  \right),\\
	\overline{I(x_m,v_m)} &= \frac{1}{2} \log \left[ \left( \frac{k_{\rm B} T}{\kappa \Delta_x^2} +  1  \right) \left( \frac{k_{\rm B} T}{m \Delta_v^2} +  1  \right) \right],
\end{align} 
where we have averaged the results with respect to the marginal probabilities $P(x_m)$ y $P(v_m)$, in order to obtain
a more general result. The information gained on making a measurement modifies the limit imposed by 
the second law\cite{parrondo,2daleysaga}
in terms of minimal work applied on the system as
\begin{equation}\label{2dagen}
W \geq \Delta F - \overline{I} k_{\rm B} T .
\end{equation}

%------------------------------------------------

\section{Instantaneous work}
\label{cuatro}
A limiting form of work which is useful to analyse, is that associated with an instantaneous process i.e. to change
the potential center from a position $\lambda_i =0$ to a position  $\lambda_f$ in zero time. In an experimental set up, this case is when 
the laser focus is changed instantaneously.  Here there is no exchange of heat with the environment, nor are there average changes in 
positions and velocities. Then,  Eq. \eqref{work} reduces to
\begin{eqnarray}
\mathbf{W}_{\rm Ins}& =& \frac{\kappa}{2} \left( \left( \mathbf{x}(t_f) - \lambda_f \right)^2 - \left( \mathbf{x}(0) - \lambda_i \right)^2 \right)
\nonumber \\
&+& \frac{m}{2} \left(  \mathbf{v}(t_f)^2 - \mathbf{v}(0)^2 \right).
\end{eqnarray} 
The particle begins at equilibrium at $\mathbf{x}(0) = \lambda_i$.
Since the change in position of the potential is instantaneous, the instantaneous values of position and velocity
of the particle do not change i.e. $\mathbf{x}(0)=\mathbf{x}(t_f)$ and $\mathbf{v}(0)= \mathbf{v}(t_f)$,
the resulting work is then
\begin{align}
\mathbf{W}_{\rm Ins} = & \frac{\kappa}{2}  \lambda_f^2. 
\end{align} 
this result is a consequence of the fact that on average we will find the particle in the center of the well with 
$\lambda_i=0$.

%------------------------------------------------

\section{Optimal work for prescribed displacement protocols}
\label{cinco}
In this section we will derive the minimum work protocol done on the system as we move the potential center $\lambda(t)$ from an
initial value $\lambda_i = 0$ to a fixed final value $\lambda_f$ in a finite time interval $t_f$\cite{salamon}. For this we need
the functional form for $\mathbf{v}$ that optimizes the integral in Eq. (\ref{work})
\begin{equation} \label{integral1}
f[\mathbf{v}] =  \int_0^{t_f}  \mathbf{v}(t) \int_0^t \Gamma(t-s) \mathbf{v}(s) ds  dt.
\end{equation} 
To optimize, 
we find that $\mathbf{v}$ must conform to the expression (see appendix \ref{appendix2})
\begin{equation} \label{expre1}
\int_0^{t_f} \Gamma(t-s) \mathbf{v}(s)  ds = c,
\end{equation}
where $c$ is a constant to be determined. This is a Fredholm integral of the first kind, that can be solved
by choosing the appropriate kernel. The memory effects here represent physically the inertial aspects of the dynamics. The colloidal particle remembers its mechanical state tending to conserve momentum during the characteristic time of the kernel. This consideration brings whole new regime not dominated by fluctuations and should always be borne out for sufficiently short time scales. It will be more pronounced for heavier particles or less frictional fluids.

Choosing the well known kernel\cite{ColmenaresOlivares} $\Gamma(t) = \gamma \alpha e^{-\alpha  \vert t \vert}$ where $\alpha$ fixes the memory
decay rate, the solution is given by\cite{Polya}
\begin{equation} \label{optimo}
\begin{array}{ll}
\mathbf{v} & =c \, \theta(t), \\
\mathbf{x} &= (c t + d) \, \theta(t) ,
\end{array}
\end{equation} 
where $d$ is the average initial position of the particle and $\theta(t)$ is the Heaviside function. 
Thus we have shown that the optimal protocol that minimizes the work done on the system for a fixed total displacement
is always linear in time. To find the value of $c$ that minimizes $\mathbf{W}$, we insert these expressions into 
Eq. (\ref{work}) along with the initial conditions at equilibrium $\mathbf{x}_{\rm eq}= \lambda_i$ and
$\mathbf{v}_{\rm eq} = 0$ and obtain
\begin{equation} \label{optraba}
\mathbf{W} =  \frac{\kappa}{2} \left( (c t_f + d ) - \lambda(t_f) \right)^2 + \frac{m}{2}  c^2  +  \gamma c^2 G(t_f), 
\end{equation} 
where
\begin{equation}
G(t_f) = 1+ \frac{{\rm e}^{-\alpha \, t_f} - 1}{\alpha \, t_f},
\end{equation} 
reflects the non-Markovian character of the system.
When $\alpha \rightarrow \infty$ we retrieve the Markovian limit $G(t_f) = 1$ (see Fig. 1).

The value of $c$ that minimizes Eq, \eqref{optraba} is
\begin{equation}\label{cgeneral}
c = \frac{\kappa \lambda_f t_f }{m + \kappa t_f^2 + 2 t_f G(t_f) \gamma },
\end{equation} 
\begin{figure}
      \includegraphics[width=7cm]{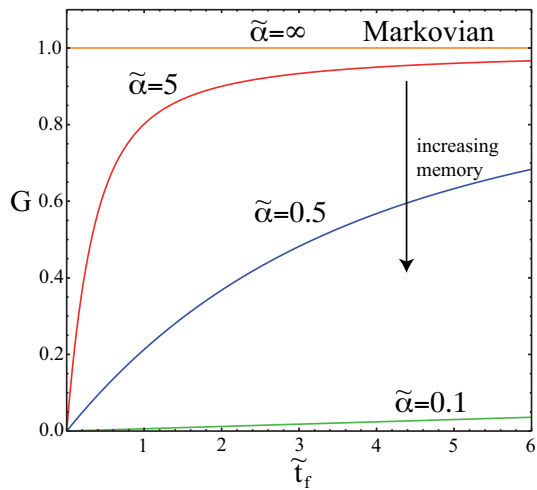}
	\label{graph1}    	
   	\caption{The inertial function $G(t_f)$ for different values of the memory parameter $\widetilde\alpha$. As $\widetilde\alpha$ increases
	we approach the Markovian limit. Note the discontinuity at $t_f=0$ when $\widetilde{\alpha}\rightarrow\infty$.} 
\end{figure}
therefore the minimum work is given by
\begin{equation}
\mathbf{W}_{\rm G} = \frac{ \kappa \lambda_f^2 }{2} - \frac{\kappa^2 t_f^2 \lambda_f^2}{2\left( \kappa t_f^2 + m + 2 t_f  \gamma G(t_f)   \right)}.
\end{equation}
where ${\rm G}$ stands for {\it general conditions} i.e. non-Markovian and underdamped.
One can derive the functional form for $\lambda(t)$, associated with the minimal work, by averaging 
Eq. (\ref{GLE}), then using
Eq. (\ref{optimo}), one obtains
\begin{align}
\label{lambdageneral}
\lambda_ {\rm G}(t) &=  \frac{\kappa \lambda_f  t_f }{m + \kappa t_f^2 + 2t_f \gamma G(t_f)} \left( t + \frac{\gamma}{\kappa}\left( 1 - e^{-\alpha t} \right) \right)   \\
&+  \frac{ m \lambda_f t_f }{m + \kappa t_f^2 + 2 t_f \gamma G(t_f)} \delta(t) \notag ,
\end{align} 
where $\delta(t)$ is the Dirac delta function to account for the sudden change in velocity at the beginning of the protocol.

In ref. \cite{GomezMarin}, the underdamped and Markovian (no memory effects) limit of this problem was addressed. They required, as part of the protocol associated with optimal work, that the velocity return to the equilibrium value at the end of the process\cite{salamon,Seifert1,GomezMarin,SeifAbreu}. The way to enforce this condition is 
to build into the protocol, a final velocity by placing an ad hoc delta function 
at the end of the process. To make contact with this limit (a sudden thermalization of the particle) we also imposed this condition, and obtained
\begin{equation} \label{likeseifert}
\mathbf{W}_{\rm S} =  \frac{\kappa}{2} \left( (c_S t_f + d ) - \lambda(t_f) \right)^2  +  \gamma c_S^2 G(t_f), 
\end{equation}
where the subindex ${\rm S}$ indicates the protocol of sudden particle thermalization at $t_f$. It should be noted that there is no mass dependent
term, because the final jump of velocity to zero eliminates any kinetic energy change.  For this protocol we have that 
\begin{equation}
c_{\rm S} = \frac{\kappa  \lambda_f }{\kappa t_f + 2 G(t_f) \gamma }.
\end{equation} 
and then
\begin{equation}
\mathbf{W}_{\rm S} =  \frac{\kappa}{2} \lambda_f^2 - \frac{ t_f \kappa^2 \lambda_f^2 }{ 2(\kappa t_f  + 2 G(t_f) \gamma)},
\end{equation} 
with the protocol
\begin{align}	
\lambda_{\rm S}(t) &=  \frac{\kappa \lambda_f  }{ \kappa t_f + 2 \gamma G(t_f)} \left( t + \frac{\gamma}{\kappa}\left( 1 - e^{-\alpha t} \right) \right)   \\
&+  \frac{ m \lambda_f }{ \kappa t_f + 2 \gamma G(t_f)} \left( \delta(t)- \delta(t_f - t) \right) \notag.
\end{align} 
In the markovian limit ($\alpha\rightarrow \infty$), we recover the results in ref. \cite{GomezMarin}.

It is useful to compare, for illustrative purposes, the latter protocol with the case where the final state of the velocity is not required
to be at equilibrium. Here, the final state of the system is out of equilibrium in general and will reach equilibrium
outside the operation of the protocol. We have not optimized again without the condition of relaxation to equilibrium at the final time
which would have yielded ${\bf W}_{\rm G}$. 
For this illustrative comparison, the protocol is given by 
\begin{align}
\lambda_{\rm N}(t) &=  \frac{\kappa  \lambda_f }{\kappa t_f + 2 G(t_f) \gamma } \left( t + \frac{\gamma}{\kappa}\left( 1 - {\rm e}^{-\alpha t} \right) \right)  \\
&+ \frac{ m  \lambda_f  }{\kappa t_f + 2 G(t_f) \gamma } \delta(t) \notag,
\end{align} 
where the label ${\rm N}$ denotes non-equilibrium final state. The corresponding optimal work $\mathbf{W}_{\rm N}$ performed is
\begin{eqnarray}
\mathbf{W}_{\rm N} &=& \frac{\kappa}{2}\lambda_f^2 +\frac{m}{2} \left( \frac{  \kappa \lambda_f  }{ 2(\kappa t_f  + 2 G(t_f) \gamma)} \right)\notag \\
  &-& \frac{ t_f \kappa^2 \lambda_f^2 }{ 2(\kappa t_f  + 2 G(t_f) \gamma)}.
\end{eqnarray} 
The difference with $\mathbf{W}_{\rm S}$ is in the second term, that in the non-equilibrated $\mathbf{W}_{\rm N}$ represents the kinetic energy that the particle acquires because of the initial velocity applied by the protocol i.e. the effect of inertia. This contribution is countered in $\mathbf{W}_{\rm S}$ by the resetting required to the equilibrium velocity at $t_f$. Obviously, these velocity contributions are not an issue in the overdamped regime because the velocity of particle is instantaneously thermalized. 

All plots for the optimal work will now be discussed in terms of the reduced variables: $\widetilde{t}_f = \kappa t_f / \gamma $,  $ \widetilde{\alpha} = \gamma \alpha / \kappa $, $\widetilde{m} = \kappa m / \gamma^2$, $\widetilde{\lambda} =  \lambda / \sqrt{k_{\rm B} T / \kappa}$, $ \widetilde{\mathbf{W}} = \mathbf{ W}/ k_{\rm B} T $. 
Note that the parameter $\widetilde{m}$ indicates the regime of the oscillator, either underdamped ($\widetilde{m} > 1/4$) or overdamped ($\widetilde{m} < 1/4$).

\begin{figure}     
     \includegraphics[width=\linewidth]{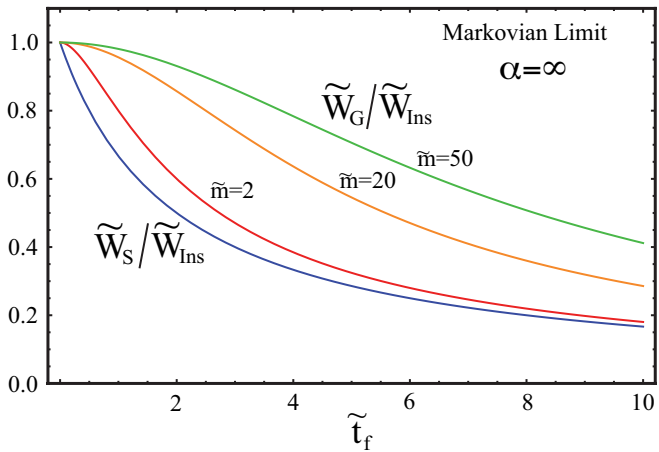}
   	\caption{Comparison between ${\bf W}_{\rm S}$ and ${\bf W}_{\rm G}$ for the Markovian limit ($\widetilde{\alpha} \rightarrow \infty$), for different
	particle mass values. As the mass increases the protocol ${\rm G}$ is less efficient and is always improved upon by the ${\rm S}$ protocol. We consider parameters $\widetilde{\lambda}_f = 2$ and work values are normalized to the ${\bf W}_{\rm Ins}$.}
\label{graph2}
\end{figure}

First, we will depict the optimal work performed in the Markovian limit, $\widetilde{\alpha} \rightarrow \infty$. Fig. \ref{graph2} shows that the smallest amount of work is performed by protocol ${\rm S}$, $\mathbf{W}_{\rm S}$. When  $t_f = 0$, $\mathbf{W}_{\rm G}$ and $\mathbf{W}_{\rm S}$ coincide and give
the instantaneous protocol value i.e only the kinetic contribution. For times $t_f >0$ the  ${\rm S}$ protocol is better (costs less work) due to the energy put into the system used to achieve the starting velocity $\lambda_{\rm G}(t)$, departing from the equilibrium value. This energy is lost at the final time because the system is left out of equilibrium (see discussion of protocol ${\rm N}$ above).  At long times the two cases coincide since friction dampens the initial states, eventually yielding the overdamped case. Note the zero slope of the work function at $t_f=0$ is indicative of inertial effects as the decay is quadratic in $t_f$ form small $t_f$. The slope of $\mathbf{W}_{\rm S}$ is $-\kappa^2\lambda_f^2/4\gamma$ so the decay of $\mathbf{W}_{\rm S}$ is linear close to $t_f=0$, that is, we can make it decay more rapidly by increasing the spring constant of the potential or decreasing the friction parameter $\gamma$. As the inertial 
effects increase (greater mass) one can see that the $G$ protocol increases its cost for the same time duration.

\begin{figure}
     \includegraphics[width=\linewidth]{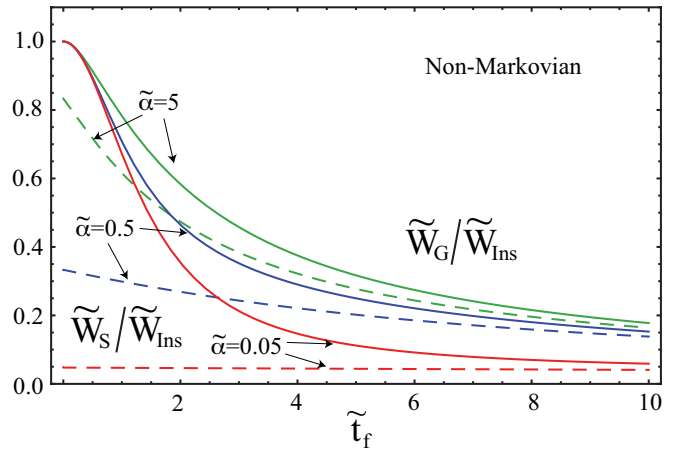}
   	\caption{Comparison between ${\bf W}_{\rm S}$ and ${\bf W}_{\rm G}$ for a range of memory parameters $\widetilde{\alpha}$. As the memory
	increases the ${\rm G}$ protocol becomes more efficient but never improves on the ${\rm S}$ protocol. Here we take $\widetilde{\lambda}_f = 2$ y $\widetilde{m} = 2$.}
     \label{graph3}
\end{figure}

The different optimal work functions show marked differences in the presence of non-Markovian effects. In Fig. \ref{graph3} we depict
the ${\rm S}$ and ${\rm G}$ protocols, normalized by the instantaneous limit, as a function of the protocol time $t_f$ and for different values of $\widetilde{\alpha}$, the
memory parameter. For short protocols $\widetilde{t}_f\rightarrow 0$, $\mathbf{W}_{\rm G}$ is the most costly due to the contribution of the velocity jump at the start of the protocol.  $\mathbf{W}_{\rm S}$ does not coincide with $\mathbf{W}_{\rm G}$ at $t_f=0$ (the difference being $\widetilde\alpha/(1+\widetilde\alpha)$) because the former does not include inertial effects (in Eq. (\ref{likeseifert}) the mass does not appear). In the end jump, the equilibrium velocity is imposed and the state of the system is reset, but the memory of the reservoir is not, so there is a time scale inconsistency at short times. 
When memory effects are more pronounced ($\alpha$ smaller) i.e. less Markovian, the ${\rm G}$ protocol enhances its efficiency but never
overtakes the ${\rm S}$ protocol.

%------------------------------------------------

\section{Optimal work for finite time protocols and information }\label{seis}
In this section we will generalize our approach to include measurements on the particle (either position or/and velocity) and thus generate 
out-of-equilibrium initial states. For the case where one measures the position of the particle, the initial distribution is given by $P_i(x,v) = P_{\rm eq}(v) P(x|x_m)$. Thus the process of measuring alters the initial conditions for the protocol by changing the distribution of positions based on the measured value. Using these initial conditions in Eq. \eqref{work} we arrive at
\begin{equation} 
\mathbf{W} =  \frac{\kappa}{2} \left( (c_x t_f + b_x ) - \lambda(t_f) \right)^2 + \frac{m}{2}  c_x^2  +  \gamma c_x^2 G(t_f).
\end{equation}
 We repeat the analysis of the previous section and find that $c_x$ now takes the value
\begin{equation}
c_x = \frac{\kappa t_f (\lambda_f - b_x)}{m + \kappa t_f^2 + 2 t_f \gamma G(t_f)}, 
\end{equation} 
where the subindex $x$ indicates a position measurement, and $b_x$ (given under Eq. (\ref{Pxm})) is averaged over the distribution $P(x|x_m)$. 
Starting from this value we compute the optimal work
\begin{equation}
\mathbf{W}_x = \frac{\kappa}{2} \lambda_f \left( \lambda_f -2 b_x  \right) - \frac{\kappa^2 t_f^2 ( \lambda_f - b_x )^2  }{2 ( m + \kappa t_f^2 +2 \gamma t_f G(t_f)  )},
\end{equation} 
and the corresponding protocol is
 \begin{eqnarray}
\lambda_x(t) &=&  \frac{ \kappa  t_f (\lambda_f - b_x) }{m + \kappa t_f^2 + 2t_f G(t_f)} \left( t + \frac{\gamma}{\kappa}\left( 1 - {\rm e}^{-\alpha t} \right) \right) \notag  \\
&+&  \frac{ m  t_f (\lambda_f - b_x) }{m + \kappa t_f^2 + 2 t_f G(t_f)} \delta(t).
\end{eqnarray}
 Averaging over the distribution $P(x_m)$ (see Eq. (\ref{measurementxm}) to obtain
\begin{equation}\label{generalwork}
\overline{ \mathbf{W}}_x = \frac{\kappa \lambda_f^2}{2}  - \frac{ \kappa^2 t_f^2 \left( \lambda_f^2 + \frac{ (k_{\rm B} T)^2 }{  \kappa ( k_{\rm B} T +  \kappa \Delta_x^2   )   } \right)  }{2 ( m + \kappa t_f^2 +2 \gamma t_f G(t_f)  )}.
\end{equation} 
According to Eq. (\ref{2dagen}), the lower limit of  $\overline{\mathbf{W}_x}$
is given by the difference
\begin{equation}\label{eq6x}
\overline{\Delta F} -\overline{I(x_m)} = - \frac{1}{2} \log \left( \frac{k_{\rm B} T}{\kappa  \Delta_x^2} +  1  \right).
\end{equation} 
Nevertheless, the limit value for the optimal work in the quasi-static limit from Eq. (\ref{generalwork}) is
\begin{equation}\label{work46}
\lim_{t_f \rightarrow \infty} \overline{\mathbf{W}}_x  = -\frac{k_{\rm B} T}{2} \frac{1}{\frac{ \kappa \Delta_x^2}{k_{\bf B} T} +  1}.
\end{equation}

If we could convert all information into work, the limits of Eq. (\ref{eq6x}) and Eq. (\ref{work46}) should be the same. 
Their difference shows the impossibility of taking advantage of all the information obtained from the measurement even with the optimal protocol.
In ref. \cite{SeifAbreu}, Abreu and Seifert showed, in the overdamped case, that it was necessary to manipulate both $\lambda$ and $\kappa$ together in order to take advantage of all information.

\begin{figure}
     \includegraphics[width=8cm]{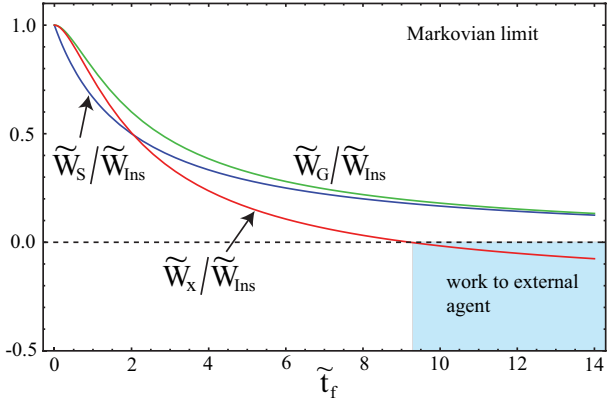}
   	\caption{Comparison of work performed for the Markovian system ($\widetilde{\alpha} \rightarrow \infty$), using the optimal protocols ${\rm G}$, ${\rm S}$ compared
	to the optimal after measuring position $x$. While $\mathbf{W}_{\rm S}$ and $\mathbf{W}_{\rm G}$ have asymptotes at zero work, $\mathbf{W}_x$ can
	return work to the external agent (shaded region). $\mathbf{W}_x$ becomes more efficient than any of the non-measuring protocol at a threshold value of time. Here the parameters take on the values: $\widetilde{\Delta}^2_x= 0.2$, $\widetilde{\lambda}_f = 2$ y $\widetilde{m} = 2$  }
     \label{graph5} 
\end{figure}

\begin{figure}
     \includegraphics[width=8cm]{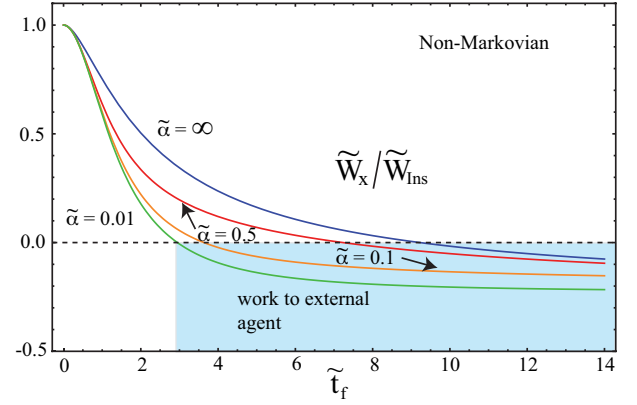}
   	\caption{Comparison between $\overline{\mathbf{W}}_x$ for different values of the memory parameter $\widetilde{\alpha}$. As the
	memory increases the protocol becomes more efficient in extracting work. We selected the
	values: $\widetilde{\Delta}^2_x= 0.2$, $\widetilde{\lambda}_f = 2$ and $\widetilde{m} = 2$ }
     \label{graph6} 
\end{figure}

In Fig. \ref{graph5} we compare the result in Eq. (\ref{generalwork}) with $\mathbf{W}_{\rm G}$ and  $\mathbf{W}_{\rm S}$ in the Markovian limit. In the underdamped regime we see that the three protocols depart from the same point, and even though $\mathbf{W}_{\rm S}$ is the minimum of the three, $\mathbf{W}_x$ becomes the optimal after a time $\widetilde{t}^*_f = \sqrt{1 + {\widetilde m} \widetilde{\lambda}_f^2(1+ \widetilde{\Delta}_x^2)} -1 $, where $\widetilde{\Delta}_x^2=\kappa {\Delta}_x^2/k_{\rm B} T$. 
This crossing is due to short time inertial dynamics of the particle that beyond $\widetilde{t}^*_f$, due to memory effects, turns into a faster reduction of the work performed by the external agent.  
Fig. \ref{graph6} depicts the effect of increasing memory effects (decreasing $\alpha$), where we observe that $\mathbf{W}_x$ decays more rapidly as $\alpha$ decreases, indicating that the protocol takes advantage of the reservoir's memory. For $t_f \rightarrow \infty$, $\mathbf{W}_x$ ceases to
depend on $\alpha$, as the reservoirs memory is erased.

For measurements of velocity, the distribution after the measurement is given by $P_i(x,v)=P_{\rm eq}(x) P(v|v_m)$. Performing 
the analysis for this distribution we express the work  in Eq. \eqref{work} as
\begin{eqnarray} 
\mathbf{W}_v &=&  \frac{\kappa}{2} \left( (c_v t_f + d ) - \lambda(t_f) \right)^2 + \frac{m}{2} \left( c_v^2 + b_v^2\right)\nonumber\\
&+&  \gamma c_v^2 G(t_f), 
\end{eqnarray} 
where $b_v$ is averaged over the distribution $P(v|v_m)$. We find that the value of  $c_v$ is the same as that found in Eq. (\ref{cgeneral})
(with no measurements), and thus the protocol is the same. Nevertheless, the optimal work function
$\mathbf{W}_v$ is indeed different
\begin{equation}
\mathbf{W}_v =  \frac{\kappa \lambda_f^2 }{2}  -  \frac{ \kappa^2 t_f^2  \lambda_f^2  }{2 ( m + \kappa t_f^2 +2 \gamma t_f G(t_f)  )} - \frac{b_v^2}{2}.
\end{equation} 
 Averaging with respect to $P(v_m)$ one arrives at
\begin{eqnarray}
\overline{\mathbf{W}}_v &=&  \frac{\kappa  \lambda_f^2 }{2}  - \frac{ \kappa^2 t_f^2  \lambda_f^2  }{2 ( m + \kappa t_f^2 +2 \gamma t_f G(t_f)  )}\notag \\
 &-& \frac{(k_{\rm B} T)^2}{2(k_{\rm B} T + m \Delta_v^2)}.
\end{eqnarray} 
One obtains different work functions departing from the same protocol $\lambda(t)$  (Eq. (\ref{lambdageneral})), 
due to the fact that the measurement changes the initial state. The final time-independent term yields an
intrinsic advantage to measuring the velocity over measuring the position that can be seen as an downward offset at $t_f=0$.

The behavior at long times, when the velocity is measured, is given by
\begin{equation}
\lim_{t_f \rightarrow \infty} \overline{\mathbf{W}}_v = -\frac{k_{\rm B} T}{2} \frac{1}{\frac{ m \Delta_v^2}{k_{\rm B} T} +  1}.
\end{equation} 
For this case, Eq. (\ref{2dagen}) gives
\begin{equation}\label{eq6v}
\Delta \overline{F} -\overline{I(v_m)} = - \frac{1}{2} \log \left( \frac{k_{\rm B} T}{m \Delta_v^2} +  1  \right). 
\end{equation} 
which is again different from the work $\overline{\mathbf{W}}_v$ in the long time limit, so that the information gained is not all made available to do work.

\begin{figure}
     \label{graph7} 
     \includegraphics[width=8cm]{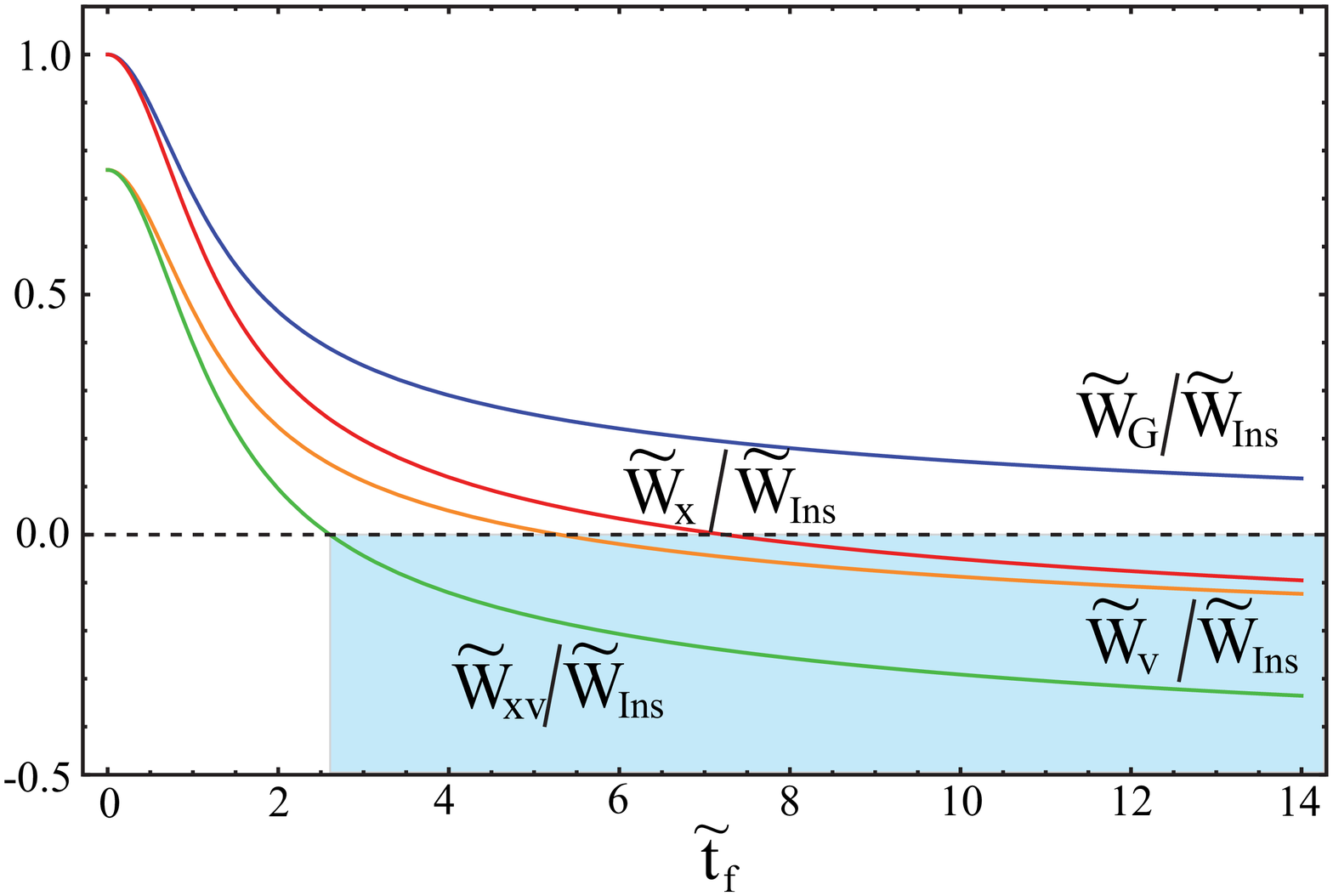}
   	\caption{Comparison of the work performed for the protocols $\mathbf{ W}$, $\overline{\mathbf{W}}_x$, $\overline{\mathbf{W}}_v$ and $\overline{\mathbf{W}}_{xv}$ for the non-Markovian evolution ($\widetilde{\alpha} =0.5 $). All measured protocols are able to return work to the
	external agent. The information derived from the velocity is more advantageous than that of the position. The parameter values used are $\widetilde{\Delta}^2_x= 0.2$, $\widetilde{\Delta}^2_v= 0.2$, $\widetilde{\lambda}_f = 2$ and $\widetilde{m} = 2$.}
\end{figure}
When one performs a simultaneous measurement of position and velocity, the initial distribution is given by $P_i(x,v) = P(x|x_m) P(v|v_m)$.
The work according to Eq. \eqref{work} is given by
\begin{equation}
\begin{aligned} 
\mathbf{W}_{xv} &=  \frac{\kappa}{2} \left( (c_{xv} t_f + b_x ) - \lambda(t_f) \right)^2 + \frac{m}{2} \left( c_{xv}^2 + b_v^2 \right)\\  &+  \gamma c_{xv}^2 G(t_f).
\end{aligned} 
\end{equation} 
In this case we see that the result is a combination of the previous cases: we find that $c_{xv} = c_x$
and that the work is modified in the same way as when we measure the velocity. This way we see that the work $\mathbf{W}_{xv}$ averaged
with respect to the probabilities $P(x_m)$ and $P(v_m)$ has the form 
\begin{eqnarray}
\overline{\mathbf{W}}_{xv} &=&  \frac{\kappa \lambda_f^2}{2}  -  \frac{ \kappa^2 t_f^2 \left( \lambda_f^2 +\frac{ (k_{\rm B} T)^2 }{ \kappa (k_{\rm B} T + \Delta_x^2 \kappa   )   } \right)  }{2 ( m + \kappa t_f^2 +2 \gamma t_f G(t_f)  )}\notag \\
 &-& \frac{(k_{\rm B} T)^2}{2(k_{\rm B} T + m \Delta_v^2)},
\end{eqnarray} 
depicted in Fig. 6. Note that the work with velocity measurements does not agree with the instantaneous work because the measurement
has an effect similar to that of a forcing a condition on the system. We see also that more information is recovered implies more work extracted
from the particle. In the long time limit 
\begin{equation}
\lim_{t_f \rightarrow \infty} \overline{\mathbf{W}}_{xv} =-\frac{k_{\rm B} T}{2} \left( \frac{1}{\frac{ \kappa \Delta_x^2}{k_{\rm B} T} +  1} + \frac{1}{\frac{ m \Delta_v^2}{k_{\rm B} T} +  1} \right) ,
\end{equation}
this quantity is smaller than that in either position (Eq. (\ref{eq6x})) and velocity  (Eq. (\ref{eq6v})) measurements. Nevertheless, for simultaneous
measurements the Sagawa relation given by Eq. (\ref{2dagen}) yields 
\begin{eqnarray}
\Delta \overline{F} -\overline{I(x_m,v_m)} &=& - \frac{1}{2} \log \left[  \left( \frac{k_{\rm B} T}{\kappa \Delta_x^2} +  1  \right)\right.
\nonumber\\
&\times&\left.\left( \!\frac{k_{\rm B} T}{m \Delta_v^2}+ 1  \right)\right],
\end{eqnarray} 
still smaller than can be achieved from manipulating the center of the well in the quasi-static limit.
This is to be expected since the measurement leads to a factorizable result
each of which cannot take advantage of the full information attained. 

\begin{figure}[b]
\label{grafnuevo}
\includegraphics[width =8cm]{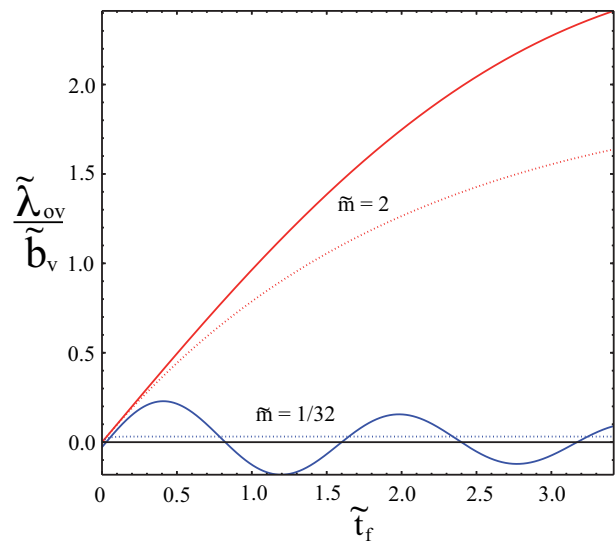}
\caption{ $\lambda_{ov}$ plots for the parameters in the legend. The dotted lines show the Markovian limit, while the solid lines are for
$\widetilde{\alpha}= 0.5$.}
\end{figure}

%------------------------------------------------

\section{Zero work protocol with velocity measurement} \label{siete}
In this section we will show that it is possible to concoct a non-optimal work protocol that can improve on optimal measurement protocols within a range of times.

We propose to do this by setting $\mathbf{v}(0)= b_v$, after a measurement of the velocity, and imposing the condition $\mathbf{x}(t)-\lambda_{ov}(t)=c$, so that the particle will see a constant potential. The velocity of the particle will be given by 
Eq. (\ref{GLE}) averaged with respect to $P(x,v,t)$
\begin{equation} \label{free}
m  \dot{\mathbf{v}} + \int_0^t \Gamma(t-s) \mathbf{v}(s) ds + \kappa c= 0,
\end{equation} 
where the value of $c$ is given by the initial condition. If the particle is at equilibrium then $\mathbf{x}(0) = \lambda_i$ so that $c=0$.

It is easy to show, for these conditions, that $\overline{\mathbf{W}}_{ov}=0$ for all values of $t_f$, so this protocol is better at short times,
(before the work done on the system becomes negative). To obtain the functional form of $\lambda_{ov}(t)$ one should solve 
Eq. (\ref{free}) using the Laplace transform
\begin{equation}
\hat{\mathbf{v}}(s) = b_v \frac{m}{m s + \hat{\Gamma}} ,\nonumber
\end{equation}
for the kernel $\Gamma(t) = \gamma \alpha {\rm e}^{-\alpha  \vert t \vert}$. One obtains the expression

\begin{eqnarray}
\lambda_{ov}(t) &=& \frac{b_v e^{-\alpha t /2}}{\omega \gamma} \left[ \left( \gamma - \frac{m \alpha}{2} \right) \left( \sinh\big[ \omega t \big]\right.\right. \nonumber\\
&-& \left.\left. m\omega \cosh\big[ \omega t\big] \right) +  m \omega e^{\alpha t/2}    \right] + \mathbf{x}(0).
\end{eqnarray}
where $\omega=\sqrt{\alpha \left( \alpha/4 - \gamma/m\ \right)}$.
We observe that this protocol does no depend  on $\lambda_f$ or $t_f$ or demand velocity jumps at the begining or end of the protocol. The
shortest $t_f$ for which $\overline{\mathbf{W}}_{xv}=0$ is reached in the limit $\alpha \rightarrow 0$ and is given by
\begin{equation}
t_f^* = \sqrt{ \frac{m \left( \kappa \lambda_f^2 - \frac{ (k_{\rm B} T)^2}{k_{\rm B} T + m \Delta^2_v}  \right)}{\kappa \frac{ (k_{\rm B} T)^2}{k_{\rm B} T + m \Delta^2_v} + \frac{ (k_{\rm B} T)^2}{k_{\rm B} T + \kappa \Delta^2_x}}  }.
\end{equation} 
This will be the longest interval for which the proposed protocol $\lambda_{ov}$ will be of benefit, for longer times the optimal
protocol $\lambda_v$ will be more efficient. In Fig. 7 we compare protocol $\lambda_{ov}$ for different regimes.
We note that both in the underdamped as well as in the overdamped regime, the systems with memory allow to attain longer $\lambda_f$ that
again give special benefits to protocols measuring the velocity.

%------------------------------------------------

\section{Summary and Conclusions}

For the paradigm of a colloidal particle bound in a harmonic potential, we have studied how to extract work controlling the center of the potential 
$\lambda(t)$. We contemplate memory effects i.e. non Markovian properties, in the underdamped regime and the measurement of 
position and velocity. We first derive the general result that optimal work protocols with and without measurements of position and velocity are shown to be linear in time, for an exponential memory kernel, as in the Markovian case. When dealing with the underdamped and non-Markovian regime one must address forcing the particle at the start and at the end of a protocols, since the velocities are not instantaneously relaxed by the reservoir. Such forcing conditions dominate the short time dynamics of the colloidal particle. 

For protocols without measurement of the position or velocity, the least work by an external agent decreases linearly
for forced start-stop conditions while those only forced at starting conditions are quadratic (slower to reduce work by agent) at short times, 
while both decrease asymptotically to zero work for quasi processes.

When measurements are performed, protocols with start-end forcing are still more efficient at short times but can be
overtaken by start-forced protocols at a threshold time. It is only for measurement protocols that one can extract
work from the particle, for long enough times. Nevertheless, the work derived is always below the maximum
predicted by Sagawa's generalization of the second law. Velocity measurement protocols are more efficient in deriving work than position measurements, and simultaneous measurements from equilibrium states have additive properties in the quasi-static limit.

Finally we derived a non-optimal protocol that uses velocity measurements to perform zero work for the
short time dynamics, thus surpassing optimal protocols until the latter reach the time at which work can be derived from the system.

As far as we know there are no works in the literature addressing optimal protocols in non-Markovian systems with inertia as the one posed in this manuscript.  However, it is relevant to mention work on the inertial Markovian Langevin equation by Gomez-Marin {\it et al.}\cite{GomezMarin} which was used as the seminal procedure to treat our problem. As we have shown, our results coincide with theirs at vanishing decay rate $\alpha$ of the colored noise. This approach does not exclude other methodologies to attack this problem without resorting to the GLE. In fact, Sivak and Crooks\cite{sivak} derived optimal protocols by calculating the time variation of work due to external perturbations through an analysis of the metric distance, thermodynamic length\cite{crooks3}, between equilibrium states. They found a similar protocol to that of Gomez-Marin et al. without considering the ad hoc velocity discrete delta jumps at the beginning and end, because the intrinsic velocity of protocols in their description change smoothly at the boundaries. We assume that our non-Markovian results should be consistent with this approach and add more information about the behavior of the system. 

Analytical treatment of non-harmonic profiles could be treated by the following strategy: First derive an appropriate fluctuation-dissipation relation for the
static external potential applied in order to arrive at the appropriate GLE. The Generalized Fokker-Planck Equation follows (subject to analytical tractability) from which the procedures in this paper can be used. This is already  feasible fot the case treated here of a particle confined in a harmonic potential but more general relations have been obtained for smoothly varying external potential which can include anharmonicities\cite{col1}. The authors in Ref.\cite{ColmenaresOlivares} have also derived fluctuation-dissipation theorems when there is a time dependent potential

%------------------------------------------------
\acknowledgments
We thank Mayra Peralta and Claudio Chamon for useful discussions. We also thank Boston University Department
of Physics who hosted one of us E.M. where this work was started. PJC thanks Universidad de Los Andes (Venezuela)
for support through Grant CDCHT-CVI-ADG-C09-95.

%------------------------------------------------

\appendix
%------------------------------------------------

\section{The work functional}
To derive the work functional that we use we will start from the expression for the heat in Eq. (\ref{HeatProduced}) substituting the
form for the probability currents given in Eq. (\ref{FPEG})
\begin{align}
\frac{\left<d Q\right> }{dt}  &=  \frac{  k_{\rm B} T }{m}  \int_0^t \Gamma(t-s) \frac{d \chi_v(t-s)}{dt} ds\\
& -  \left< v(t) \int_0^t \Gamma(t-s) v(s) ds \right>.\notag 
\end{align} 
Substituting into Eq. (\ref{Work}) together with the expression for the energy we have
\begin{align*}
\left< W \right> & = \frac{m}{2} \left< v(t_f)^2 \right> + \frac{k}{m} \left< \left( x(t_f) - \lambda(t_f) \right)^2 \right> \\
& - \frac{m}{2} \left< v(0)^2 \right> - \frac{k}{m} \left< \left( x(0) - \lambda(0) \right)^2 \right>\\ 
& - \int_0^{t_f} \left[ \int_0^t \frac{  k_{\rm B} T }{m}  \Gamma(t-s) \frac{d \chi_v(t-s)}{dt} ds \right. \\
& \left. -  \left< v(t) \int_0^t \Gamma(t-s) v(s) ds \right>  \right] dt.
\end{align*} 
Rewriting in terms of variances and average values one can write
\begin{align*}
\left< W \right> &= \frac{m}{2} \left( \sigma^2_v(t_f) - \sigma^2_v(0)  \right) + \frac{k}{2}\left( \sigma^2_x(t_f) - \sigma^2_x(0)  \right) \\
&+ \frac{k}{2} \left[\left( \left< x(t_f) \right> - \lambda(t_f) \right)^2 - \left( \left< x(0) \right> - \lambda(0) \right)^2   \right] \\ 
& + \frac{m}{2}\left( \left<v(t_f)\right>^2 - \left<v(0)\right>^2  \right)\\
& - \int_0^{t_f} \left[ \int_0^t \frac{  k_{\rm B} T }{m}  \Gamma(t-s) \frac{d \chi_v(t-s)}{dt} ds \right.\\
& \left. -  \left< v(t) \int_0^t \Gamma(t-s) v(s) ds \right>  \right] dt.
\end{align*}
To simplify this expression we resort to Eq. (\ref{FPEG}) from which we can derive the system of dynamical equations
for the first moments of the positions and velocities
\begin{align*}
\frac{d \left< x \right>}{d t} &= \left< v \right>,\\
\frac{d \left< v \right>}{dt} &= -\left< \int_0^t \frac{\Gamma(t-s)}{m} v(s) ds \right> \\
&- \frac{k}{m} \left( \left< x \right> - \lambda(t) \right),
\end{align*}
and for their second moments
\begin{eqnarray}
\frac{d \left< x^2 \right>}{dt} &=& 2 \left< xv \right>,\nonumber\\
\frac{d \left< v^2 \right>}{dt} &=& -2 \left< v(t) \int_0^t \frac{\Gamma(t-s)}{m} v(s) ds \right> \nonumber\\
&+ &2 \frac{\kappa}{m} \left( \lambda(t) \left< v \right> - \left< xv \right> \right)\nonumber\\ 
& +& \frac{ k_{\rm B} T }{m} \int_0^t \Gamma(t-s) \frac{d \chi_v(t-s)}{dt}ds,\nonumber\\
\frac{d \left< vx \right>}{dt} &=& \left< v^2 \right> - \left< x(t) \int_0^t \frac{\Gamma(t-s)}{m} v(s) ds \right>\nonumber \\
& +&  \frac{k}{m} \left( \lambda(t) \left< x \right> - \left< x^2 \right> \right)\nonumber\\
& +& \frac{ k_{\rm B} T }{m} \int_0^t  \Gamma(t-s)  \chi_v(t-s)ds.
\nonumber
\end{eqnarray}
Using the previous expressions we can write
\begin{eqnarray}
&&\frac{k}{m}\frac{d \sigma^2_x(t)}{d t}+\frac{d \sigma^2_v(t)}{d t}  + 2 \left<  v(t) \int_0^t \frac{\Gamma(t-s)}{m} v(s) ds \right>
\nonumber\\
& -2& \frac{ k_{\rm B} T }{m} \int_0^t \Gamma(t-s) \frac{d \chi_v(t-s)}{dt} ds\nonumber\\
&=& 2 \left<  v(t) \right> \left< \int_0^t \frac{\Gamma(t-s)}{m} v(s) ds \right>.\nonumber
\end{eqnarray} 
Integrating with respect to $t$ from $0$ to $t_f$ one finds
\begin{align*}
&\int_0^{t_f}\!\!\!\! \left<  v(t) \right> \left< \int_0^t \Gamma(t-s) v(s) ds \right> dt  =  \frac{m}{2} \left( \sigma^2_v(t_f) - \sigma^2_v(0)  \right) \\
&+ \frac{k}{2}\left( \sigma^2_x(t_f)\! \!- \sigma^2_x(0)  \right) \!-\!\! \int_0^{t_f}\!\! \left[ \int_0^t \!\!\frac{ k_{\rm B} T }{m} \Gamma(t-s) \frac{d \chi_v(t-s)}{dt} ds \right. \\
 & \left.-  \left< v(t) \int_0^t \Gamma(t-s) v(s) ds \right>  \right] dt,
\end{align*} 
which we can use to reduce the work function to the form
\begin{eqnarray}
\left< W \right> &=& \frac{m}{2}\left( \left<v(t_f)\right>^2 - \left<v(0)\right>^2  \right) + \frac{k}{2} \left[\left( \left< x(t_f) \right> - \lambda(t_f) \right)^2 \right.\nonumber  \\
 &-& \left.  \left( \left< x(0) \right> - \lambda(0) \right)^2 \right]\nonumber\\
 & +& \int_0^{t_f} \left<  v(t) \right> \left< \int_0^t \Gamma(t-s) v(s) ds \right> dt . \notag
\end{eqnarray}
Once we recognize the notation where bold fonts represent averages over $P(x,v,t)$ we obtain Eq. (\ref{work}).

%------------------------------------------------

\section{Optimization procedure}
\label{appendix2}
Given the functional
\begin{equation} \label{integral2}
f[\mathbf{v}] =  \int_0^{t_f}  \mathbf{v}(t) \int_0^t \Gamma(t-s) \mathbf{v}(s) ds  dt.
\end{equation} 
which can be more generally written as 
\begin{equation}
f[\mathbf{x}]=\int_0^{t_f}  \mathcal{F}(t,\mathbf x,\mathbf v) dt,
\end{equation}
then the functional derivative $\delta f[\mathbf x]$ is given by
\begin{equation}
\delta f[ \mathbf{x},h(t)] = \left [\frac{d f}{d \varepsilon}\right ]_{\varepsilon=0}=\int_0^{t_f} \frac{\delta \mathcal{F}(t,\mathbf x,\mathbf v)}{\delta \mathbf x} h(t) dt, 
\end{equation} 
where $h$ is the variation of $\mathbf x$, an auxiliary function and $\dot{\mathbf{x}} = \mathbf{v}$. The functional derivative can be expressed 
as
\begin{equation}
\begin{aligned}
\delta f[\mathbf{x},h(t)] &=\left [\frac{d}{d\varepsilon} \int_0^{t_f} \mathcal{F}(t,\mathbf x+\varepsilon h(t),\dot{\mathbf x}+\varepsilon \dot h(t))dt\right ]_{\varepsilon=0} \\
&= \int_0^{t_f} \left[ \int_0^t \dot{h}(t) \Gamma(t-s) \dot{\mathbf{x}}(s) ds \right. \\ 
&+ \left. \int_0^t \dot{\mathbf{x}}(t) \Gamma(t-s) \dot{h}(s)  ds  \right] dt.
\end{aligned}
\end{equation}
Interchanging the integral limits for the second term, we obtain
\begin{eqnarray}
\delta f[\mathbf{x},h] &=& \int_0^{t_f} \left[ \int_0^t \dot{h}(t) \Gamma(t-s) \dot{\mathbf{x}}(s) ds \right. \nonumber\\ 
&+& \left. \int_t^{t_f} \dot{\mathbf{x}}(s) \Gamma(s-t) \dot{h}(t)  ds  \right] dt.
\end{eqnarray} 
Taking advantage of the parity of the kernel, one can simplify it to
\begin{equation}
\delta f[\mathbf{x},h] = \int_0^{t_f} \dot{h}(t) \int_0^{t_f} \Gamma(t-s) \dot{\mathbf{x}}(s) ds dt,
\end{equation} 
and then integrate by parts to obtain
\begin{equation}
\delta f[\mathbf{x},h] = \int_0^{t_f} h(t) \frac{d}{dt} \left( \int_0^{t_f} \Gamma(t-s) \dot{\mathbf{x}}(s) ds \right) dt.
\end{equation} 
To optimize, the last equation must be set to zero, hence we find that $\mathbf{v}$ must conform to the expression
\begin{equation} \label{expre2}
\int_0^{t_f} \Gamma(t-s) \mathbf{v}(s)  ds = c.
\end{equation}

%------------------------------------------------

%------------------------------------------------

\end{document}